\documentclass[journal]{IEEEtran}
\ifCLASSINFOpdf
\else
\fi

\usepackage{amsmath}
\usepackage{amsfonts}
\usepackage{xcolor}
\usepackage{graphicx}

\begin{document}
\title{Multimodal Feature Fusion and Knowledge-Driven Learning via Experts Consult for Thyroid Nodule Classification}

\author{Danilo Avola,~\IEEEmembership{Member,~IEEE,}
        Luigi Cinque,~\IEEEmembership{Senior Member,~IEEE,}
        Alessio Fagioli,~\IEEEmembership{Student Member,~IEEE,}
        Sebastiano Filetti, Giorgio Grani, 
        Emanuele Rodolà,~\IEEEmembership{Member,~IEEE}% <-this % stops a space
\thanks{D. Avola, L. Cinque, A. Fagioli, and E. Rodolà are with the Department of Computer Science, Sapienza University, Via Salaria 113, 00198, Rome, Italy (e-mails: \{avola,~cinque,~fagioli,~rodola\}@di.uniroma1.it)}% <-this % stops a space
\thanks{S. Filetti and G. Grani are with the Department of Translational and Precision Medicine, Policlinico Umberto I, Sapienza University, Viale Regina Elena 328, 00161, Rome, Italy. (e-mails: \{sebastiano.filetti, giorgio.grani\}@.uniroma1.it)}% <-this % stops a space
\thanks{Manuscript received November 10, 2020; revised ?}}


\markboth{IEEE TRANSACTIONS ON CIRCUITS AND SYSTEMS FOR VIDEO TECHNOLOGY, VOL. XX, NO. XX, MONTH XXXX}%
%{Avola \MakeLowercase{\textit{et al.}}: Multimodal Feature Fusion and Knowledge-Driven Learning via Experts Consult for Thyroid Nodule Classification}
{Multimodal Feature Fusion and Knowledge-Driven Learning via Experts Consult for Thyroid Nodule Classification}



\maketitle

% \IEEEpubid{\begin{minipage}{\textwidth}\ \\[12pt] \centering
% Copyright \copyright 2021 IEEE. Personal use of this material is permitted. However, permission to use this material for any other purposes must be obtained from the IEEE by sending an email to pubs-permissions@ieee.org.
% \end{minipage}}

% \IEEEpubidadjcol

\begin{abstract}
Computer-aided diagnosis (CAD) is becoming a prominent approach to assist clinicians spanning across multiple fields. These automated systems take advantage of various computer vision (CV) procedures, as well as artificial intelligence (AI) techniques, to formulate a diagnosis of a given image, e.g., computed tomography and ultrasound. Advances in both areas (CV and AI) are enabling ever increasing performances of CAD systems, which can ultimately avoid performing invasive procedures such as fine-needle aspiration. In this study,
a novel end-to-end knowledge-driven classification 
\color{black}
framework is presented. The system focuses on multimodal data generated by thyroid ultrasonography, and acts as a CAD system by providing a thyroid nodule classification into the benign and malignant categories. Specifically,
\color{black}
the proposed system leverages cues provided by an ensemble of experts to guide the learning phase of a densely connected convolutional network (DenseNet). The ensemble is composed by various networks pretrained on ImageNet, including AlexNet, ResNet, VGG, and others. The previously computed multimodal feature parameters are used to create ultrasonography domain experts via transfer learning, decreasing, moreover, the number of samples required for training. To validate the proposed method, extensive experiments were performed, providing detailed performances for both the experts ensemble and the knowledge-driven DenseNet. As demonstrated by the results, the proposed system achieves relevant performances in terms of qualitative metrics for the thyroid nodule classification task, thus resulting in a great asset when formulating a diagnosis.
\end{abstract}


% \fi
\IEEEpeerreviewmaketitle

\section{Introduction}
\IEEEPARstart{T}{hyroid} nodules, described by an abnormal growth of the gland tissue, are a common disease affecting the thyroid gland \cite{moon2011ultrasonography}. Ultrasonography is the most used modality to both detect and diagnose nodules. This method is safe, convenient, non-invasive, and has a better capability of distinguishing benign nodules from malignant ones, with respect to other techniques such as computed tomography (CT) and magnetic resonance imaging (MRI), thus facilitating early diagnosis and treatment choice \cite{schlumberger2015lenvatinib}. In order to take full advantage of ultrasound (US) images, computer-aided diagnosis (CAD) is rapidly evolving, resulting in systems able to provide less subjective interpretations and, consequently, more precise diagnoses. A CAD system is generally developed following established phases including image preprocessing (e.g., noise removal, image reconstruction), region-of-interest (ROI) extraction, segmentation, and classification. Historically, many of the available works focus on the first three steps, while in the latest years the emphasis is being shifted towards thyroid nodule classification due to the evolution of machine learning approaches. A key aspect of all phases lies in the nodule representation, where techniques such as local derivative and local binary patterns (LDP, LBP) in \cite{zhang2009lbp}, or discrete wavelet transform (DWT) in \cite{shensa1992dwt}, can provide a detailed description of the gland itself. By applying these techniques, as well as a plethora of other computer vision approaches, it is ultimately becoming possible to detect and segment the thyroid inside a US image \cite{chang2010thyroidsegandvolestim,zhao2013segmentation}, or classify its nodules \cite{grani2020contemporary}, thus making CAD systems a great asset for clinicians during their diagnoses. 
\IEEEpubidadjcol

\begin{figure*}[t]
    \centering
    \includegraphics[width=\textwidth]{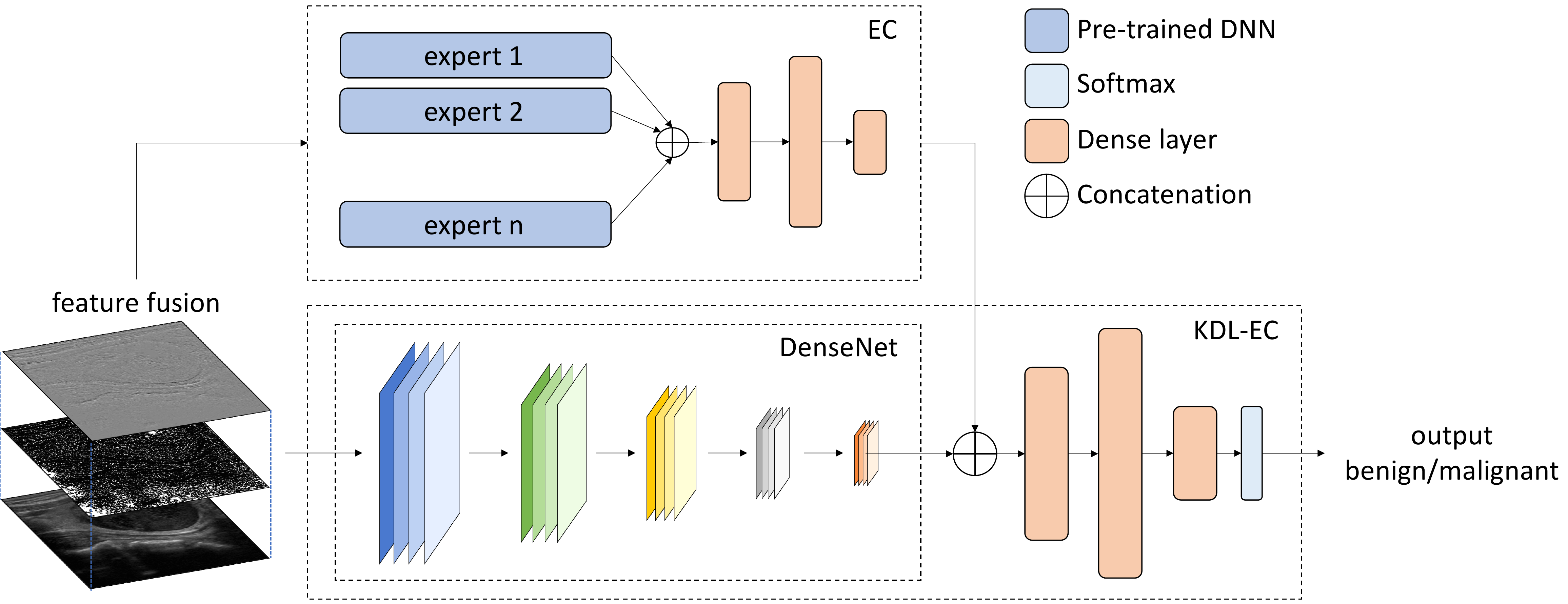}
    \caption{Knowledge-driven learning (KDL) via experts consult (EC) framework architecture. The feature fusion between US, LBP and DWT images is given as input to the EC module and to the KDL-EC DenseNet. Their outputs are then concatenated and elaborated by three dense layers to obtain a diagnosis. The EC module, through an ensemble of deep neural networks (DNN), can guide the KDL-EC unit learning.}
    \label{archi}
\end{figure*}

%\section{Related Work}
In the latest years, many interesting works using various machine learning approaches concerning the nodule classification task, were presented. The authors of \cite{wu2016classifier}, for example, propose a comparison between Bayesian techniques, support vector machines (SVMs), and neural networks (NNs), where the latter show promising results which are close to classic radiological methods. Similarly, by using the expectation–maximization (EM) algorithm to train a convolutional neural network (CNN), a neural network able to grasp correlations in an image through convolutions, in \cite{wang2018learning}, the performances of an automatic system are further improved by training the system in a semi-supervised way. 
While convolutional networks are certainly powerful, representing the nodules in a meaningful way can give an edge on the diagnosis accuracy, especially when few samples are available. In \cite{liu2017classification}, histogram of oriented gradients (HOG) and LBP descriptors are successfully combined with other high-level features, to compensate the lack of thyroid images.
A different approach to solve the issue of small datasets, common problem for medical researches, is data augmentation. For this procedure, traditional methods include image cropping, rotation, and scaling, although even neural networks, such as generative adversarial networks (GANs), can be used to increase the dataset size. Indeed, the authors of \cite{zhu2017image} perform data augmentation through CNNs, to achieve improved performances with respect to standard US. Another approach to handle small-sized datasets, is transfer learning. Through this method, a network trained on a different task is reused to reach convergence faster and more easily on the new domain. In \cite{elasticityfusion}, this technique is exploited in combination with a feature fusion procedure, where US images of thyroid glands are fused together with their respective elasticity maps, ultimately exceeding state-of-the-art performances.

While much was done concerning the classification through neural networks and transfer learning, other approaches are also emerging based on either ensemble learning or domain knowledge transfer. By using an ensemble, for example, radiologists performances for thyroid nodule classification are exceeded in \cite{li2019diagnosis}. Other works successfully applying and, thus, encouraging the use of the ensemble technique are devised in \cite{gupta2017fibernet} and \cite{xie2017transferable}. The former is able to perform white matter fiber clustering, while the latter can classify lung nodules from CT scans with high accuracy. Regarding the domain knowledge transfer, knowledge itself can be used to effectively drive the learning phase of a network via the prior knowledge of a different one \cite{avola2019master}. As a matter of fact, the authors of \cite{liu2018integrate} apply a similar knowledge-driven rationale to diagnose breast cancer by integrating domain knowledge during the training phase, ultimately showing that this approach (i.e., domain knowledge transfer) can be used to formulate a medical diagnosis.

In this study, due to the relatively small dataset at our disposal, we apply both data augmentation and feature fusion on automatically generated multimodal data, to fully exploit the available US images. Moreover, inspired by the results obtained in \cite{avola2019master}, we explore a novel knowledge-driven learning (KDL) approach by building an ensemble of experts via transfer learning, and using it to guide the learning phase of another network, i.e., acting as an experts consult (EC). Indeed, the ensemble provides consults on the input thyroid image and can effectively guide another network, i.e., a DenseNet in this work, during its training. As a result, the knowledge-driven DenseNet can converge faster, requires a lower training time, and obtains a higher accuracy with respect to both the ensemble or a simple DenseNet learning the task by itself.
This outcome is also supported by the experimental results, suggesting that a knowledge-driven learning via experts consult (KDL-EC) approach can correctly guide a network during its training, even on complex medical images. Indeed, as shown in the experiments, while a direct comparison cannot be performed due to each state-of-the-art work using a different private dataset, the proposed system obtains relevant accuracy, precision, recall, f1, and AUC scores in comparison with the other available solutions, when analysing the same type of images (i.e., thyroid US). Thus, the presented method results in a great performing thyroid nodule classifier that can be used as a CAD system.

Summarizing, key contributions can be resumed as follows:
\begin{itemize}
    \item for the first time in the state-of-the-art, the presentation of a multimodal feature fusion approach based on raw ultrasound (US) images, local binary pattern (LBP) and discrete wavelet transform (DWT) representations; 
    \item for the first time in the literature, the use of knowledge-driven learning via experts consult (KDL-EC) approach to directly improve the training phase of a densely connected convolutional neural network (DenseNet), directly applied on medical images;
    \item the introduction of an end-to-end framework comprising a data augmentation generation phase (via LBP and DWT), an ensemble of experts fine-tuning (EC module), and the description of a novel knowledge-driven approach (KDL-EC unit).
\end{itemize}

The remaining part of this paper is organized as follows. Section 2 introduces the proposed knowledge-driven learning via experts consult framework. Section 3 analyses both quantitatively and qualitatively the method performances, and presents a comparison with other literature works. Finally, Section 4 draws some conclusions on the presented framework.

\section{Materials and methods}
The proposed knowledge-driven learning via experts consult framework, shown in Figure~\ref{archi}, can be divided into three components: a data augmentation and multimodal feature fusion phase, where detail-rich nodule images are generated; an expert consult (EC) module based on the ensemble stacking technique, where pre-trained deep neural networks are fine-tuned; and a knowledge-driven learning (KDL) unit, where EC cues on the most likely diagnosis are used to guide a standalone convolutional network during its training.

%%%%%%%%%%%%%%%%%%%%%%%%%%%%%%%%%%
%%%%%%%%%%%%%%%%%%%%%%%%%%%%%%%%%%
%% DATA AUGMENTATION AND FUSION %% ref Fig2: Figure~\ref{nodulesfusion}
%%%%%%%%%%%%%%%%%%%%%%%%%%%%%%%%%%
%%%%%%%%%%%%%%%%%%%%%%%%%%%%%%%%%%
\subsection{Data augmentation and multimodal feature fusion}
\begin{figure}[t]
    \centering
    \includegraphics[width=\columnwidth]{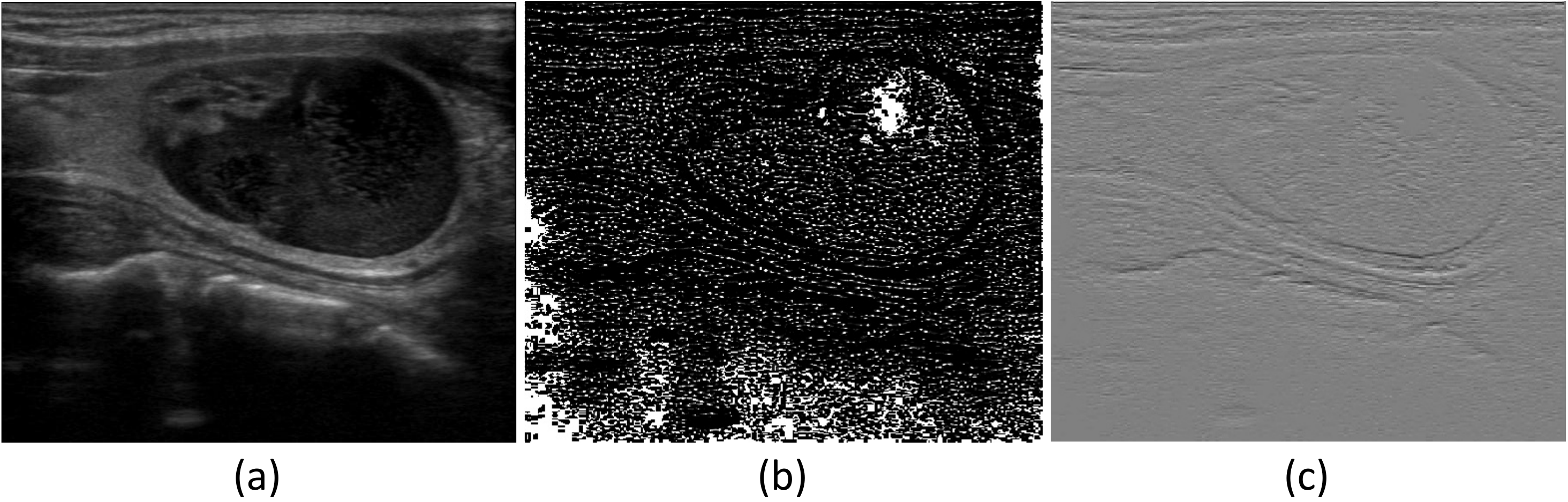}
    \caption{Data augmented nodule example. In (a), the raw US image, while (b) and (c) show the same nodule analysed via LBP and DWT, respectively.} \label{nodulesfusion}
\end{figure}

The first augmentation and fusion component is a mandatory choice for the proposed framework, due to the used dataset being relatively small. 
\color{black}
In this work, differently from common strategies where, for example, geometrical transformations or GANs are employed to produce new samples \cite{shorten2019survey}, data augmentation is performed by applying the LBP and DWT algorithms to each US thyroid nodule.
An original multimodal feature fusion is then obtained by stacking the nodule US with its corresponding LBP and DWT images, along their channels axis. Since all images are grayscale, this procedure results in an object representing a nodule with shape $(w*h*3)$, where $w$ and $h$ correspond to width and height of the image; while each channel is a different representation of the nodule. LBP and DWT representations were chosen since they provide further information about both inner and outer properties of the nodule itself, as shown in Figure~\ref{nodulesfusion}. What is more, these two algorithms (i.e., LBP and DWT), have already been used, although in standalone solutions, with a similar rationale in other relevant medical works such as breast cancer classification \cite{wan2017integrated}, colonic polyps analysis \cite{wimmer2016directional}, as well as many others \cite{huang2020segmentation,pepe2020detection,stolte2020survey}.
Finally, the feature fusion object is used as input for the other two components.

\begin{figure}[t]
	\centering
    \includegraphics[width=\linewidth]{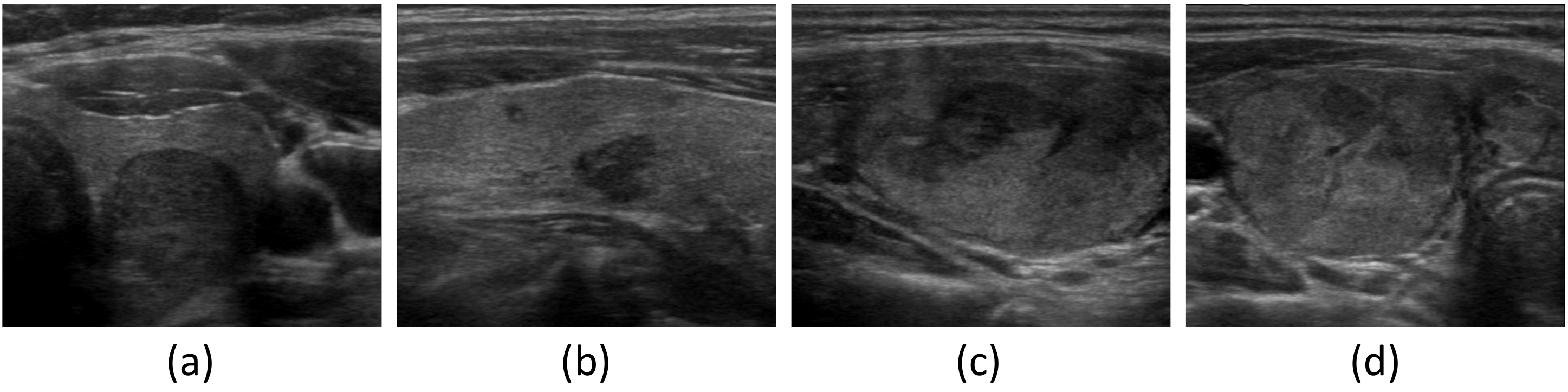}
	\caption{Images from the devised dataset. Benign examples are shown in (a) and (b), while malignant nodules are represented in (c) and (d).} 
	\label{nodules}
\end{figure}

\begin{table*}[t]
    \centering
    \caption{Average 10-fold cross-validation models accuracies of baseline pre-trained models and various percentages of frozen layers. In each configuration block, accuracies on raw US, augemented (i.e., US, LBP and DWT) and feature fusion datasets (i.e., left, center, and right column of each block, respectively), are shown.}
    \label{finetuningaccshort}
    %\resizebox{0.8\textwidth}{!}{
    \begin{tabular}{l|c|c|c|c|c|c|c|c|c|c|c|c}
        \hline
        
        Model & 
        \multicolumn{3}{c|}{Baseline\%} & 
        \multicolumn{3}{c|}{Frozen-25\%} & 
        \multicolumn{3}{c|}{Frozen-50\%} & 
        \multicolumn{3}{c}{Frozen-75\%} \\ 
        %\hline
        \hline
        
        AlexNet & 70.2 & 71.4 & 74.2 &
        75.5 & 76.7 & 78.1 &
        79.9 & 82.0 & \textbf{83.1} &
        73.8 & 74.0 & 74.8 \\
        %\hline
        
        DenseNet & 
        76.8 & 78.0 & 81.5 &
        80.6 & 82.8 & \textbf{87.2} &
        78.3 & 79.6 & 83.4 &
        77.4 & 78.5 & 82.0 \\
        %\hline
        
        GoogleNet & 
        71.8 & 73.2 & 76.9 &
        80.0 & 82.3 & \textbf{84.9} &
        74.9 & 77.8 & 81.6 &
        72.0 & 76.5 & 79.3 \\
        %\hline
        
        ResNet & 
        74.5 & 75.3 & 77.4 &
        76.5 & 79.3 & 81.2 &
        78.7 & 81.2 & \textbf{84.6} &
        75.7 & 76.7 & 78.7 \\
        %\hline
        
        ResNeXt & 
        73.8 & 74.3 & 77.9 &
        76.7 & 80.5 & 82.5 &
        82.0 & 83.1 & \textbf{84.7} &
        76.2 & 77.5 & 80.7 \\
        %\hline
        
        VGG & 
        75.6 & 76.4 & 80.3 &
        77.3 & 81.5 & 83.1 &
        81.2 & 83.4 & \textbf{85.6} &
        76.8 & 78.9 & 81.6 \\
        \hline
    \end{tabular}
    %}
\end{table*}

%%%%%%%%%%%%%%%%%%%%
%%%%%%%%%%%%%%%%%%%%
%% EXPERT CONSULT %%
%%%%%%%%%%%%%%%%%%%%
%%%%%%%%%%%%%%%%%%%%
\subsection{EC}
In the second component, i.e., EC, an ensemble stacking module is fine-tuned in order to later help training another network. The ensemble is composed by $n$ pre-trained deep neural networks, defined experts in this work, taking as input the described feature fusion object, which is normalized via a convolutional layer to handle the different thyroid representations. Each expert is first modified to output a binary classification by changing the last dense layer size of a given network to 2 units. The modified expert is then fine-tuned on the thyroid dataset. An ensemble is built using these fine-tuned experts, so that all members can operate simultaneously to formulate a single diagnosis.
To obtain a stacking ensemble, predictions of all $n$ experts are concatenated and re-elaborated through 3 dense layers, so that the opinion (i.e., prediction) of each expert is taken into account for the final diagnosis.
The first two dense layers utilize a rectified linear unit (ReLU) activation function and effectively merge together all of the $n$ experts predictions. The third layer employs a softmax function to obtain a probability distribution over the two available classes (i.e., benign and malignant), which is then used to produce a diagnosis on the multimodal feature fusion input object, representing a thyroid nodule.
During this re-elaboration phase, the new dense layers weights are modified while all the fine-tuned experts are left untouched, i.e., the training error does not back-propagate beyond the 3 dense layers used to implement the stacking technique.
As experts, several good performing networks pre-trained on ImageNet \cite{deng2009imagenet} were used, namely: AlexNet \cite{krizhevsky2012imagenet}, DenseNet \cite{huang2017densely}, GoogleNet \cite{szegedy2015going}, ResNet \cite{he2016deep}, ResNeXt \cite{xie2017aggregated}, and VGG \cite{simonyan2014very}. These pre-trained networks allow to apply the transfer learning technique, where previous knowledge is transferred and used on a new domain. Notice that while the chosen models are trained on non-medical images, some common characteristics, such as object contours, are still present in the new domain and can be effectively used on the new task, as already shown by \cite{elasticityfusion}. Through transfer learning, it is possible to reduce both time and number of samples required to fully train a network, thus making this technique ideal for the proposed ensemble which is based on hard-to-come-by medical images. Finally, once the ensemble stacking module is trained on the thyroid images, its diagnoses are used to provide a medical consult and guide the learning of the last module of the proposed framework.

%%%%%%%%%%%%%%%%%%%%%%%%%%%%%%%%%%%%%
%%%%%%%%%%%%%%%%%%%%%%%%%%%%%%%%%%%%%
%% KNOWLEDGE-DRIVEN LEARNING (KDL) %%
%%%%%%%%%%%%%%%%%%%%%%%%%%%%%%%%%%%%%
%%%%%%%%%%%%%%%%%%%%%%%%%%%%%%%%%%%%%
\subsection{KDL-EC}
The third and last framework component, is the KDL-EC unit. Intuitively, in this module, an external ensemble (i.e., the EC) provides meaningful cues to a standalone convolutional neural network so that it can achieve better performances on its classification task.
In this study, a DenseNet is chosen as CNN since, due to its dense connections between convolutions, it is able to forward propagate relevant information during training. Moreover, this network has already obtained interesting results on diverse tasks analysing both medical \cite{savardi2018beta} and non-medical images \cite{avola20192d,xie2020multiscale}. 
Similarly to the external EC module, the standalone KDL-EC unit receives as input a multimodal feature fusion object, which is normalized via a convolutional layer so that the DenseNet can fully exploit the different thyroid representations (i.e., raw US, LBP, and DWT images). The DenseNet output is then concatenated to the EC cues and re-elaborated via 3 dense layers to produce a diagnosis prediction, comparably to the ensemble stacking technique. The key aspect of the KDL-EC unit lies in the training error which, unlike the EC module, is back-propagated through the whole DenseNet during the training phase, thus enabling the novel knowledge-driven approach. Moreover, all possible prediction errors are exclusively attributed to the DenseNet itself. Indeed, the meaningful EC cues are exploited to formulate a diagnosis prediction (i.e., via concatenation), although the training error is not back-propagated to the external standalone EC.

Formally, to build the KDL-EC unit, the DenseNet is extended so that its output is concatenated to the EC module prediction and re-elaborated via three dense layers. Similarly to the EC unit, the first two dense layers make use of the ReLU activation function, while the last one utilizes a softmax function to obtain a probability distribution over the two available classes (i.e., benign and malignant). The training error, propagating only through the DenseNet, is obtained using a binary cross-entropy loss function, computed via the following equation:
\begin{equation}
    \mathcal{L}=-\left(y\log(p)+(1-y)\log(1-p) \right),
\end{equation}
where $y$ indicates the image label (i.e., 0 for benign and 1 for malignant); $p$ represents the probability computed by the softmax layer; and $\log(\cdot)$ corresponds to the natural logarithm.

\begin{figure}[t]
	\centering
	\includegraphics[width=\linewidth]{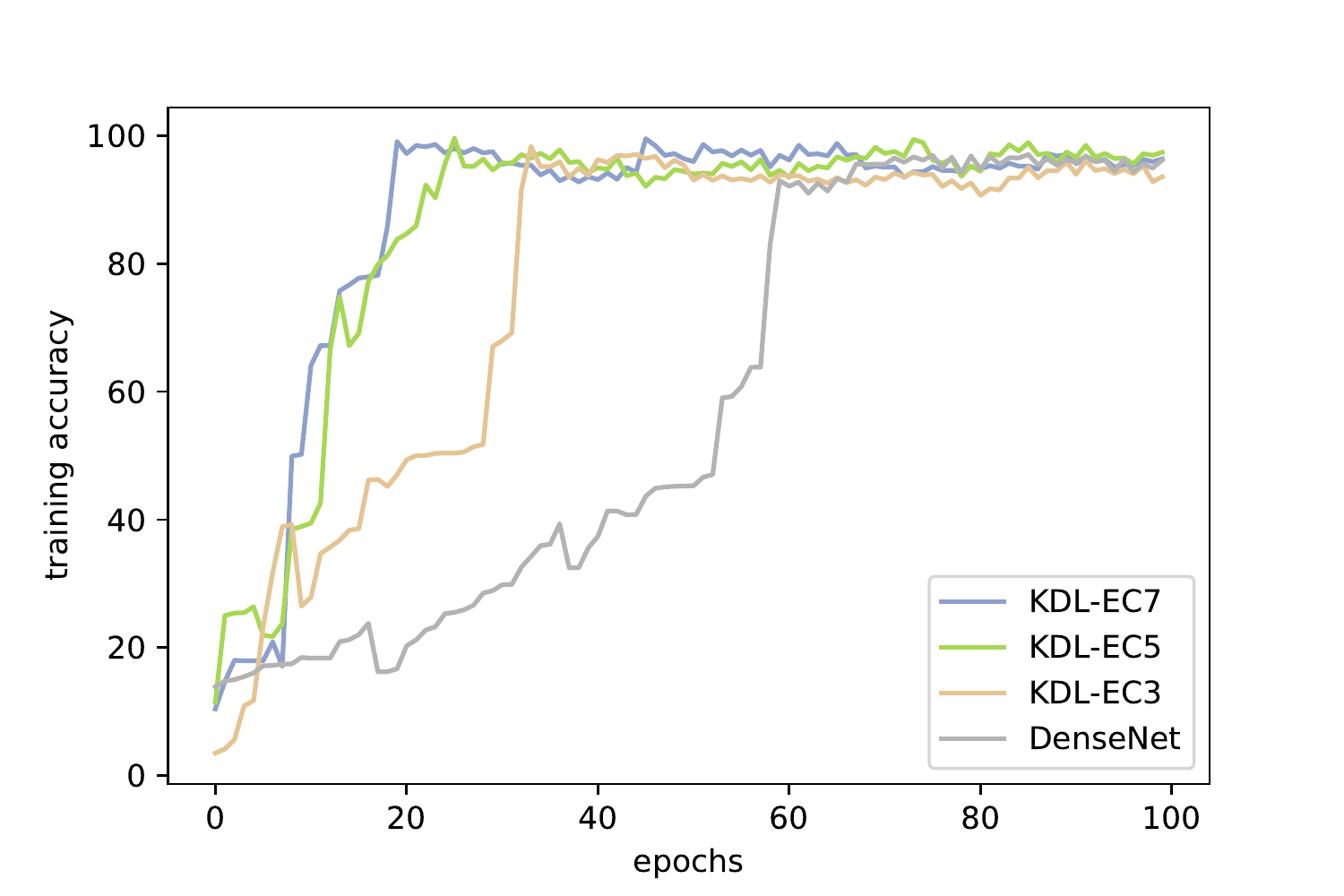}
	\caption{Average 10 fold cross validation accuracy on dataset $D2$.} 
	\label{convergence}
\end{figure}

\section{Experimental results}
In this section, the private dataset and the hardware configuration used to test the proposed framework are first introduced. A qualitative analysis on the multimodal feature fusion effectiveness, and the experimental results for all the mentioned components, are then presented and discussed.

%%%%%%%%%%%%%%%%%%%%%%%%%%%%%%%%%%%%%%%%
%%%%%%%%%%%%%%%%%%%%%%%%%%%%%%%%%%%%%%%%
%% DATASET AND HARDWARE CONFIGURATION %% ref Fig3: Figure~\ref{nodules}
%%%%%%%%%%%%%%%%%%%%%%%%%%%%%%%%%%%%%%%%
%%%%%%%%%%%%%%%%%%%%%%%%%%%%%%%%%%%%%%%%
\subsection{Dataset and Hardware Configuration} All the experimental results shown in this section were carried out on a private dataset of thyroid nodules, provided by the Department of Translational and Precision Medicine of the Policlinico Umberto I hospital of Rome. The study was conducted with Institutional Review Board approval (Sapienza University of Rome ethics committee) and written informed patient consent was obtained from all patients. 
The dataset, collected from 230 distinct patients, is composed by 678 unmarked grayscale ultrasound images generated directly from the DICOM format, and cropped to a size of $440\times440$ so that the thyroid gland is retained. Moreover, each image has a 
\color{black}
cytological report associated that uses a tiered classification system similar to TI-RADS. The report, indicating the tissue malignancy risk,
\color{black}
is utilised to split the available samples into the benign and malignant categories.
\color{black}
Specifically,
\color{black}
all images with a score~$\le2$ are labelled as the former, while the remaining samples (i.e., with a score~$\ge3$) are associated to the latter, thus defining a binary classification task.
\color{black}
Notice that tissues with a low malignancy risk (i.e., with a score~$=$~3) are still labeled as such to avoid missing any malignant occurrence. 
\color{black}
Examples of benign and malignant nodules, are shown in Figure~\ref{nodules}. After this nodule-label association, the dataset was split into two sets, $D1$ and $D2$, with non overlapping patients. $D1$ contains 452 samples, divided into 360 benign and 92 malignant cases while $D2$ comprises 226 images, partitioned into 180 benign and 46 malignant cases. This subdivision enables us to obtain unbiased results for both the EC and KLD-EC modules during their experimentation, since they are trained on independent datasets.
Concerning the hardware configuration, all tests are performed on the Google cloud platform (GCP), leveraging the pytorch framework and using a virtual machine with the following specifications: 4-Core Intel i7 2.60GHz CPU with 16GB of RAM, and a Tesla P100 GPU. 

\begin{figure*}[t]
	\centering
    \includegraphics[width=0.9\textwidth]{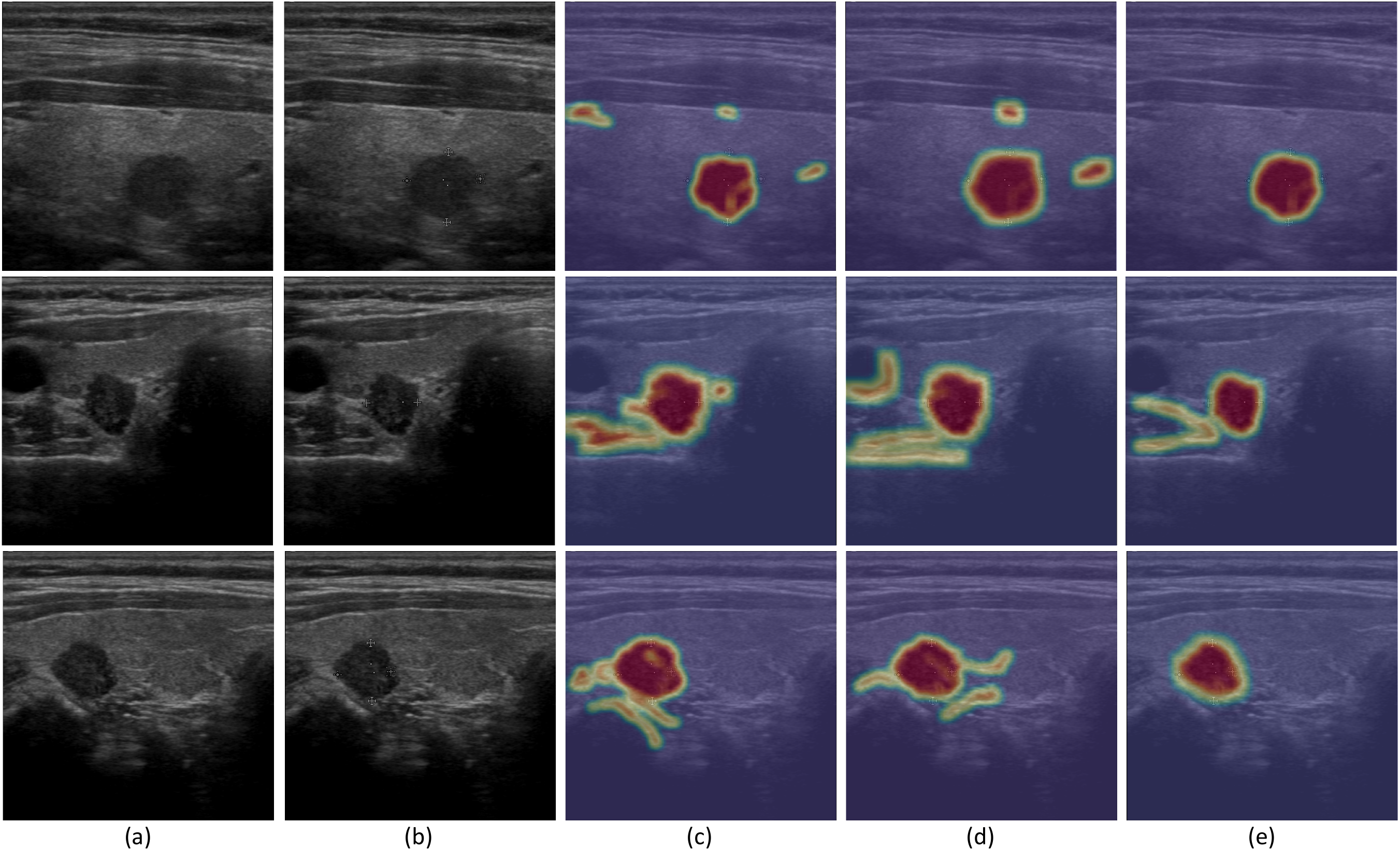}
	\caption{Grad-cam qualitative study comparison for the KDL-EC-7 DenseNet. The top row shows a benign nodule, middle and bottom rows a malignant one. In (a) and (b) the input images and their nodule marked version. In (c), (d), (e), the Grad-cam algorithm output using raw US, the augmented collection (i.e., raw US, LBP, and DWT representations as separate images), and the feature fusion dataset, respectively.} 
	\label{qualitative}
\end{figure*}

%%%%%%%%%%%%% ref Fig4: Figure~\ref{convergence}
%%%%%%%%%%%%% ref Tab1: Table~\ref{finetuningaccshort}
%% RESULTS $$ ref Tab2: Table~\ref{ecblockperf}
%%%%%%%%%%%%% ref Tab3: Table~\ref{kdlecblockperf}
%%%%%%%%%%%%% ref Tab4: Table~\ref{perfcomparison}
\subsection{Results}
In order to fully assess the proposed KDL-EC framework, a preliminary grid-search is employed to evaluate various ImageNet pre-trained networks. The networks of choice are: AlexNet, DenseNet, GoogleNet, ResNet, ResNeXt, and VGG. Each network is available with pre-trained weights in the torchvision library of the pytorch framework. For each model, the best ImageNet performing version is selected, following the scores reported in \cite{torchvisionmodels}. All networks are trained for 1000 epochs, using the early stopping technique \cite{prechelt1998automatic} to avoid overfitting, a learning rate of 0.001, a batch size of 32, and $[0.2, 1]$ as class weights to handle the discrepancy in the number of samples between the benign and malignant classes, respectively. Relevant results on the $D1$ collection, using either raw US images, augmented (i.e., US, LBP, and DWT) or multimodal feature fusion datasets, are summarized in Table~\ref{finetuningaccshort}. The shown performances were obtained using a stratified 10-fold cross-validation with random 80/20\% fold-splits and non overlapping patients between training and test sets (i.e., for a given fold, all images associated to a person are used in either of the two sets). This procedure was selected to both ensure the sample class distribution was maintained when creating train and test sets, as well as to guarantee their sample independence; thus resulting in a more accurate evaluation.
As shown, using either the augmented or multimodal fusion datasets, results in consistently improved performances. The rationale behind this behaviour can be attributed to the extra information both the LBP and DWT can provide, in the augmented collection, and the ability to directly correlate visual cues among the various representations, in the multimodal feature fusion dataset. 
Moreover, by maintaining pre-trained weights via frozen layers, it is possible to achieve conspicuous performances boosts through the transfer learning and network fine-tuning techniques, as also \cite{elasticityfusion} already demonstrated in their work. 
Indeed, as shown in Table~\ref{finetuningaccshort}, the best results for all networks are obtained by using both the feature fusion dataset and by freezing up to the first 25\% or 50\% layers of the corresponding ImageNet pre-trained network architecture. 
Notice that these values were empirically chosen in accordance with the findings of \cite{elasticityfusion}, where the best results were obtained by freezing roughly up to the first 33\% layers of their architecture; an outcome that can be directly correlated to the early layers ability to represent characteristics commonly shared among different images types such as, for example, object contours.

Concerning the experts consult unit performances, the best performing networks were chosen to create an ensemble for the EC module, according to the preliminary grid-search results. 
The EC members were selected according to Table~\ref{finetuningaccshort} scores, starting from the best performing model and in decreasing order across the various architectures. Notice that, even though the same model can be used multiple times in the consult, using networks with different internal structures allows to simulate experts with different diagnostic abilities and can result in more heterogeneous opinions (i.e., the models tend to agree less on different inputs).
All tests were conducted on the same 10-fold feature fusion dataset $D1$ used to fine-tune the single ensemble experts, to evaluate EC modules comprising 3, 5, and 7 experts.
In the EC-7, two different implementations of a DenseNet (i.e., DenseNet169 and DenseNet201 from \cite{torchvisionmodels}) were used to build the ensemble. This decision was taken due to the DenseNet obtaining the best performance on the feature fusion dataset. In Table~\ref{ecblockperf}, common evaluation metrics for a baseline network and the three EC models, are summarized. As shown, increasing the number of experts, allows the module to obtain better overall performances, even though the single networks cannot perform as well as the EC module. The rationale behind this behaviour can be attributed to the ensemble stacking technique, where the outputs of the single networks (i.e., the experts) are re-elaborated in order to produce a better representation of the input and, consequently, a more accurate output. Notice that specificity scores are slightly lower than sensitivity ones, due to the dataset being skewed toward the benign class, even though class weights are utilised to represent the difference in the number of samples.

\begin{table}[t]
    \centering
    \caption{Average 10-fold cross validation performances for the experts consult (EC) module at different sizes. The baseline scores refer to the best performing network: a feature fusion fine-tuned DenseNet. Scores are computed on 
    \color{black}
    on feature fusion
    \color{black} dataset $D1$.}\label{ecblockperf}
    \resizebox{\linewidth}{!}{
        \begin{tabular}{l|c|c|c|c}
            \hline
            Model & Accuracy\% & Sensitivity\% & Specificity\% & AUC\% \\
            \hline
            Baseline & 87.20 $\pm$ 0.77 & 87.50 $\pm$ 1.13 & 83.33 $\pm$ 1.17 & 89.12 $\pm$ 1.45 \\
            EC-3 & 89.36 $\pm$ 0.93 & 89.61 $\pm$ 0.92 & 84.95 $\pm$ 0.77 & 91.33 $\pm$ 1.11 \\
            EC-5 & 90.09 $\pm$ 0.68 & 90.37 $\pm$ 0.85 & 85.39 $\pm$ 1.02 & 92.78 $\pm$ 1.09 \\
            EC-7 & 91.25 $\pm$ 0.71 & 91.94 $\pm$ 0.73 & 86.22 $\pm$ 0.59 & 93.06 $\pm$ 1.34 \\
            \hline
        \end{tabular}
    }
\end{table}
\begin{table}[t]
\color{black}
    \centering
    \caption{Average 10-fold cross validation performances for the knowledge-driven learning via experts consult module (KDL-EC) at different sizes. The baseline scores refer to the best performing EC module. Scores are computed on feature fusion dataset $D1$.}\label{kdlecblockperf_d1}
    \resizebox{\linewidth}{!}{
        \begin{tabular}{l|c|c|c|c}
            \hline
            Model & Accuracy\% & Sensitivity\% & Specificity\% & AUC\% \\
            \hline
            %\hline
            Baseline & 91.25 $\pm$ 0.71 & 91.94 $\pm$ 0.73 & 86.22 $\pm$ 0.59 & 93.06 $\pm$ 1.34  \\
            KDL-EC-3 & 95.01 $\pm$ 0.87 & 95.99 $\pm$ 1.03 & 89.32 $\pm$ 0.70 & 96.71 $\pm$ 1.07  \\
            KDL-EC-5 & 95.27 $\pm$ 0.86 & 96.07 $\pm$ 1.21 & 90.01 $\pm$ 1.06 & 97.23 $\pm$ 1.09  \\
            KDL-EC-7 & 96.11 $\pm$ 0.95 & 96.42 $\pm$ 0.99 & 91.02 $\pm$ 1.04 & 98.03 $\pm$ 1.11  \\
            \hline
        \end{tabular}
    }
\end{table}
\begin{table}[t]
    \centering
    \caption{Average 10-cross validation performances for the knowledge-driven learning via experts consult module (KDL-EC) at different sizes. The baseline scores refer to the best performing EC module. Scores are computed on feature fusion dataset $D2$.}\label{kdlecblockperf}
    \resizebox{\linewidth}{!}{
        \begin{tabular}{l|c|c|c|c}
            \hline
            Model & Accuracy\% & Sensitivity\% & Specificity\% & AUC\% \\
            \hline
            Baseline & 91.07 $\pm$ 1.04 & 91.47 $\pm$ 1.17 & 90.09 $\pm$ 1.33 & 94.02 $\pm$ 1.27  \\
            KDL-EC-3 & 94.83 $\pm$ 0.83 & 95.78 $\pm$ 1.12 & 92.22 $\pm$ 1.13 & 97.75 $\pm$ 1.10  \\
            KDL-EC-5 & 94.95 $\pm$ 0.94 & 95.94 $\pm$ 1.03 & 92.67 $\pm$ 0.99 & 97.97 $\pm$ 1.12  \\
            KDL-EC-7 & 95.11 $\pm$ 0.99 & 96.22 $\pm$ 1.23 & 93.09 $\pm$ 1.38 & 98.79 $\pm$ 0.93  \\
            \hline
        \end{tabular}
    }
\end{table}

In relation to the KDL-EC unit performances,
\color{black}
experiments were carried out on both feature fusion datasets $D1$ and $D2$. Notice that while both collections allow to correctly evaluate the KDL-EC unit, which presents, as a matter of fact, the same behaviour on both sets, completely unbiased results are produced from trials on dataset $D2$ since the EC component was fine-tuned on dataset $D1$. Concerning the evaluation, similarly to the aforementioned test, a stratified 10-fold cross-validation, with random 80/20\% fold-splits and non overlapping patients between training and test sets, was used to evaluate this component.
Moreover, the KDL-EC unit was tested using three different experts consults composed by 3, 5, and 7 members. 
The experts ensemble networks were the same used for the EC module, and remained frozen during the training phase of the KDL-EC unit. The base network of choice for the KDL-EC was a DenseNet, selected due to obtaining the best scores in the preliminary results. 
The DenseNet was trained for 1000 epochs, using the early stopping technique to avoid overfitting, a learning rate of 0.001, a batch size of 32, and class weights set to $[0.2, 1]$ for benign and malignant samples, respectively.
Common evaluation metrics for a baseline network and the three KDL-EC models on feature fusion datasets $D1$ and $D2$ are compared in Table~\ref{kdlecblockperf_d1} and Table~\ref{kdlecblockperf}, respectively. As shown, for both collections, adding external cues based on previous knowledge during the learning phase (i.e., experts consult output), can drastically increase the performance of the DenseNet. Even more interesting, is the increase in the specificity score, which is related to the malignant samples. In this case, even though the dataset is skewed toward the benign class, the network is still able to increase its performance by leveraging the EC model output.
What is more, a knowledge-driven DenseNet is able to converge much faster compared to an unaided one. Indeed, as shown in Figure~\ref{convergence}, all KDL-EC units converge much faster (i.e., epochs 20, 26, and 34) than a single network (i.e., epoch 69) by using cues provided by the experts. This outcome indicates that the knowledge-driven learning via experts consult is a feasible approach to improve the learning phase of a network, while also reducing the time required to obtain relevant performances due to the lower number of epochs required to reach convergence.
To conclude the KDL-EC unit analysis, a qualitative test was also performed by using the Grad-cam \cite{selvaraju2017grad} algorithm, which provides a visual explanation for the network decision. This experiment was carried out by feeding the three different input types (i.e., raw US, augmented, and fused datasets) to the knowledge-driven DenseNet exploiting cues from 7 experts (i.e., KDL-EC-7). Input samples, their marked version, and results for this qualitative assessment, are shown in Figure~\ref{qualitative}. As can be seen, independently of the input type, the network is able to focus on the nodule, i.e., Figure~\ref{qualitative}.c, \ref{qualitative}.d, and \ref{qualitative}.e, even though a gland segmentation is not provided. Moreover, by also exploiting LBP and DWT representations via the augmented and fusion collections, the DenseNet obtains a more precise nodule representation (i.e., Figure~\ref{qualitative}.d and \ref{qualitative}.e). As a consequence, the network is able to capture possible irregularities which might help to discern between the benign and malignant classes. Furthermore, by using the presented feature fusion object, the knowledge-driven DenseNet can reduce the number of details which are otherwise analysed outside the nodule when using raw US, LBP and DWT representations separately; eventually resulting in the shown higher performances.

Finally, in Table~\ref{perfcomparison}, a comparison with other relevant works, is presented. 
Although each method is tested on a different private thyroid dataset, thus preventing a direct comparison, the reported results still allow to assess the proposed system performances since all models analyse US images of thyroid nodules. 
Indeed, the KDL-EC framework, thanks to its feature fusion and knowledge-driven learning approach, is able to achieve significant performances across all metrics; thus suggesting a CAD system could be improved via previous knowledge, provided experts are available to produce cues on the task at hand. 

\begin{table}[t]
    \centering
    \caption{State-of-the-art methods performances comparison.}
    \label{perfcomparison}
    \resizebox{\columnwidth}{!}{
        \begin{tabular}{l|c|c|c|c}
            \hline
            Method &\  Accuracy\% \  &\  Sensitivity\% \  &\  Specificity\% \  &\ \ \ \  AUC\% \ \ \ \  \\
            \hline
            %\hline
            \cite{wu2016classifier} & 84.74 & 92.31 & 76.00 & 91.03 \\
            \cite{liu2017classification} & 93.10 & 90.80 & 94.50 & 97.70  \\
            \cite{zhu2017image} & 93.75 & 93.96 & 92.68 & -  \\
            \cite{wang2018learning} & 88.25 & 90.00 & 86.50 & 92.86  \\
            \cite{li2019diagnosis} & 89.80 & 93.40 & 86.10 & 94.70  \\
            \cite{liu2019automated} & 94.90 & \textbf{97.20} & 89.10 & - \\
            \cite{wang2020automatic} & 87.32 & 84.22 & - & 90.06 \\
            \cite{elasticityfusion} & 94.70 & 92.77 & \textbf{97.96} & 98.77 \\
            \textbf{KDL-EC} & \textbf{95.11} & 96.22 & 93.09 & \textbf{98.79} \\
            \hline
        \end{tabular}
    }
\end{table}

\section{Conclusion}
In this paper, a novel knowledge-driven learning via experts consult framework for thyroid nodule classification is presented. The proposed system exploits the fusion of several image representations to better describe thyroid nodules and enhance the accuracy of all its components. Furthermore, the experimental results confirm that by leveraging the previous knowledge obtained by an ensemble of experts (i.e., a consult), it is possible to guide a new network during its training phase, speeding this process up, and ultimately obtaining improved results with respect to both the base network as well as the ensemble itself. 
As future work, more images are going to be collected and possibly released, so that a common ground for other works can be established. Moreover, further experiments on the proposed knowledge-driven approach utilising different types of input (e.g., elasticity maps), as well as a complementary module handling the TI-RADS classification in an automatic fashion, will also be considered. 

\section*{Acknowledgments}
This work was supported by the MIUR under grant “Departments of Excellence 2018–2022” of the Sapienza University Computer Science Department; the Sapienza University Small research project (RP11916B881DAA8F); the European Thyroid Association Research Grant for Clinical Science 2018; and the ERC Starting Grant no. 802554 (SPECGEO).

% or


\ifCLASSOPTIONcaptionsoff
  \newpage
\fi



%\bibliographystyle{IEEEtran}


% \end{thebibliography}

 \bibliographystyle{IEEEtran}
 \bibliography{bibliography}

\begin{IEEEbiography}[{\includegraphics[width=1in,height=1.25in,clip,keepaspectratio]{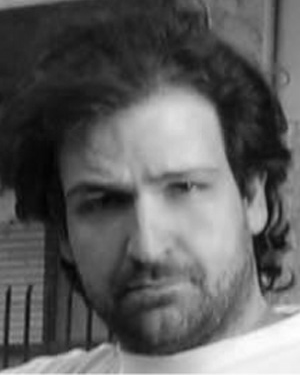}}]{Danilo Avola} earned the M.Sc. degree in Computer Science from Sapienza University, Rome, Italy, in 2002 and the Ph.D. degree in Molecular and Ultrastructural Imaging from University of L'Aquila, L'Aquila, Italy, in 2014. Since 2021 he is Assistant Professor at the Department of Computer Science of Sapienza University, where he leads both the Robotics Vision and Artificial Intelligence Laboratory (TITAN Lab) and the Computer Vison Laboratory (VisionLab). In addition, since 2018 he is R\&D Senior Engineer at the W•SENSE s.r.l., a spin-off of Sapienza University, where he leads the Computer Vision Team (CVT), and since 2010 he provides consultation and collaboration to companies engaged in computer science, computer vision, and artificial intelligence research projects. Previously, he was postdoc researcher at the Department of Mathematics, Computer Science and Physics (DMIF), University of Udine, Udine, Italy, and R\&D Senior Engineer at the Artificial Vision and Real-Time Systems Laboratory (AVIRES Lab) at the same University. Among other awards and prizes, Danilo Avola received, in 2020, the Outstanding Paper Award for IEEE Transactions on Industrial Informatics (TII). Currently, he is Reviewer of top International Journals in Computer Vision, including Pattern Recognition, Pattern Recognition Letters, IEEE Transactions on Industrial Informatics, IEEE Transactions on Industrial Electronics, IEEE Transactions on Circuits and Systems for Video Technology, IEEE Transactions on Human-Machine Systems, and International Journal of Computer Vision; in addition he is Associate Editor and Guest Editor of different ranked International Journals. His research interests include Computer Vision, Image/Video Processing, Human Computer Interaction, Wi-Fi Signal Processing, EGG Signal Processing, Machine/Deep Learning, Multimodal Systems, Pattern Recognition, Event/Action/Affect Recognition, Action, Scene Understanding, Body Language and Face Expression Interpretation, Robotics (UAVs, AUVs, ROVs, Humanoids), and has published around 100 papers on these topics. Since 2011, Danilo Avola is member of IAPR, CVPL, and IEEE.
\end{IEEEbiography}

\begin{IEEEbiography}[{\includegraphics[width=1in,height=1.25in,clip,keepaspectratio]{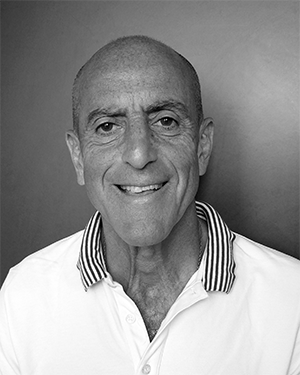}}]{Luigi Cinque} received the M.Sc. degree in physics from the University of Napoli, Naples, Italy, in 1983. From 1984 to 1990 he was with the Laboratory of Artificial Intelligence (Alenia S.p.A), working on the development of expert systems and knowledge-based vision systems. He is Full Professor of Computer Science at the Sapienza University, Rome, Italy. His first scientific interests cover Image Processing, Shape and Object Recognition, Analysis and Interpretation of Images and their Applications, with particular emphasis at Content-Based Retrieval in Visual Digital Archives, and Advanced Man-Machine Interaction assisted by Computer Vision. Currently, his main interests involve Distributed Systems for Analysis and Interpretation of Video Sequences, Target Tracking, Multisensor Data and Information Fusion. Some of techniques he has proposed have found applications in field of Video-Based Surveillance Systems, Autonomous Vehicle, Road Traffic Control, Human Behaviour Understanding, and Visual Inspection. He is author of more than 200 papers in National and International Journals, and Conference Proceedings. He currently serves as reviewer for many International Journals (e.g., IEEE Trans. on PAMI, IEEE Trans. on Circuit and Systems, IEEE Trans. on SMC, IEEE Trans. on Vehicular Technology, IEEE Trans. on Medical Imaging, Image and Vision Computing). He served an scientific committees of International Conferences (e.g., CVPR, ICME, ICPR) and symposia. He serve as a reviewer for the European Union in different research program. He is Senior member of IEEE, and member of IAPR, and CVPL.
\end{IEEEbiography}

\begin{IEEEbiography}[{\includegraphics[width=1in,height=1.25in,clip,keepaspectratio]{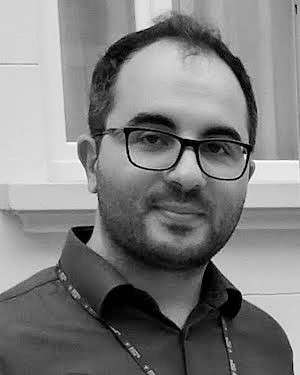}}]{Alessio Fagioli} received the B.Sc and the M.Sc. (cum laude) in Computer Science from Sapienza University of Rome, Rome, Italy, in 2016 and 2019, respectively. He is currently a Ph.D. fellow and a member of the VisionLab of the Department of Computer Science, Sapienza University of Rome. His current research interests include Medical Image Analysis, Machine Learning, Deep Learning, Affect Recognition, Event Recognition, Object Tracking, and Human Computer Interaction.
\end{IEEEbiography}

\begin{IEEEbiography}[{\includegraphics[width=1in,height=1.25in,clip,keepaspectratio]{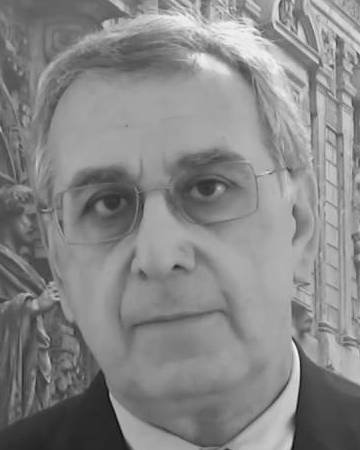}}]{Sebastiano Filetti} was a full professor of Internal Medicine at Sapienza University of Rome. Dr. Filetti is currently Editor in Chief of ENDOCRINE, Springer publishing. He is member of the American Thyroid Association since 1980 and Endocrine Society since 1983. Dr. Filetti’s current clinical interests include thyroid cancer, diabetes and endocrine/metabolic diseases. Dr. Filetti’s research interests focus on thyroid tumorigenesis and the implications of the genetic and epigenetic abnormalities in endocrine tumors. He is involved in research focusing on the identification and clinical validation of thyroid cancer biomarkers to achieve a better cancer risk stratification and on the genetic basis of familial and sporadic thyroid cancer. He is the founder of the Italian Thyroid Cancer Observatory (ITCO) register.  
\end{IEEEbiography}

\begin{IEEEbiography}[{\includegraphics[width=1in,height=1.25in,clip,keepaspectratio]{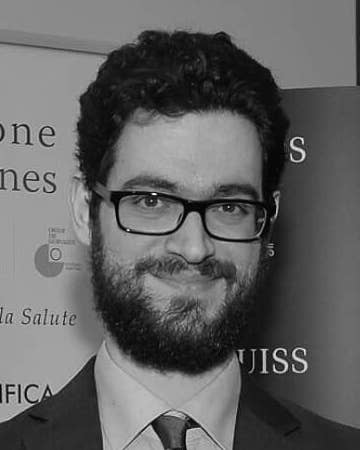}}]{Giorgio Grani} received his M.D. from the University of Rome Sapienza, where he also completed clinical training in endocrinology and the Ph.D. program in Biotechnologies in clinical medicine. His main research field is the study of thyroid nodules and cancer, particularly imaging and medical management. He regularly serves as a reviewer for international journals and currently works at the Department of Translational and Precision Medicine, Sapienza University of Rome.
\end{IEEEbiography}

\begin{IEEEbiography}[{\includegraphics[width=1in,height=1.25in,clip,keepaspectratio]{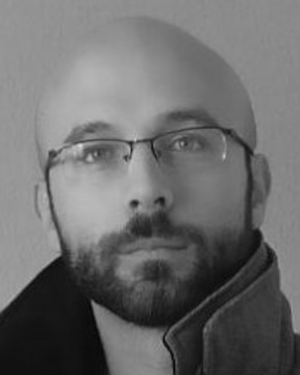}}]{Emanuele Rodol\`a} is Associate Professor of computer science at Sapienza University of Rome, where he leads the GLADIA group of Geometry, Learning \& Applied AI, funded by an ERC Starting Grant (2018--). Previously, he was a postdoc at USI Lugano, an Alexander von Humboldt Fellow at TU Munich, and a JSPS Research Fellow at The University of Tokyo. He received a number of prizes, including Best Paper Awards at 3DPVT 2010, VMV 2015, SGP 2016, 3DV 2019, and 3DV 2020. He has been serving in the program committees of the top rated conferences in computer vision and graphics (CVPR, ICCV, ECCV, SGP, etc.), as Area Chair at 3DV (2016--2019), founded and chaired several successful workshops including the workshop on Geometry Meets Deep Learning (co-located with ECCV/ICCV, 2016--2019), organized multiple SHREC contests, and was recognized 9 times as IEEE Outstanding Reviewer at CVPR/ICCV/ECCV. He has been giving tutorials and short courses on Geometric Deep Learning in multiple occasions at EUROGRAPHICS, ECCV, SGP, SIGGRAPH, and SIGGRAPH Asia. His research interests include geometry processing, geometric and graph deep learning and computer vision, and has published around 90 papers on these topics.
\end{IEEEbiography}


%\vfill


\end{document}